\documentclass[12pt,preprint]{aastex}
\usepackage{emulateapj5}
\usepackage{apjfonts}

\submitted{To Appear in the Astrophysical Journal}

\begin{document}

\title{Cluster-Cluster Microlensing as a Probe of Intracluster Stars,
MACHO\lowercase{s}, and Remnants of the First Generation Stars}

\author{Tomonori Totani\altaffilmark{1}}

\affil{Princeton University Observatory, Peyton Hall, Princeton, NJ
08544-1001, USA } 

\altaffiltext{1}{
Theory Division, National
Astronomical Observatory, Mitaka, Tokyo 181-8588, Japan }


\begin{abstract}
The galaxy cluster Abell 2152 is recently found to be forming a
cluster-cluster system with another, more distant cluster whose core is
almost perfectly aligned to that of A2152.  We discuss the detectability of
microlensing events where a single star in the source cluster behind A2152 is
extremely magnified by an intracluster compact object in A2152.  We show that
a search with an 8m-class telescope with a wide field of view, such as the
Subaru/Suprime-Cam, can probe intracluster compact objects with a wide mass
range of $m_{\rm co} \sim 10^{-5}$--$10^{10} M_\odot$, including ranges that
have not yet been constrained by any past observations.  We expect that the
event rate is biased for the background cluster than the foreground cluster
(A2152), which would be a unique signature of microlensing, making this
experiment particularly powerful.  The sensitivity of this experiment for the
mass fraction of compact objects would be 1--10\% in the total dark matter of
the cluster, which is roughly constant against $m_{\rm co}$, with a
reasonable telescope time for large telescopes ($\sim$ 10 nights). Therefore
any compact objects in this mass range can be detected or rejected as the
dominant component of the dark matter. About 10 events are expected if 20\%
of the cluster mass is in a form of compact objects with $M \sim 1 M_\odot$,
as claimed by the MACHO collaboration for the Milky Way halo.  Other possibly
detectable targets include intracluster stars stripped by galaxy
interactions, and hypothetical very massive black holes ($M \gtrsim 100
M_\odot$) produced as remnants of the first generation stars, which might be
responsible for the recently reported excess of the cosmic infrared
background radiation that seems impossible to explain by normal galactic
light.
\end{abstract}


\section{Introduction}
Gravitational microlensing of stars in the Magellanic Clouds (MCs) provides
us a unique probe of compact objects that might be a significant part of the
dark matter in the Galactic halo (Paczy\'nski 1986), and intensive effort has
been made so far (see, e.g., Narayan \& Bartelmann 1995 for a review). The
MACHO collaboration interprets the microlensing events towards MCs as
providing evidence for massive compact halo objects (MACHOs) with a mass of
$\sim 0.1$--1$M_\odot$ in the Galactic halo, constituting a significant
fraction ($\sim 20\%$) of the total halo mass (Alcock et al.  2000).  The
EROS collaboration, on the other hand, has used their observations to place
an upper limit of $\sim$10\% on the MACHO mass fraction (Lasserre et
al. 2000).

When the impact parameter is much smaller than the Einstein radius, a very
strong magnification is expected. By using such strongly magnified events,
often called pixel lensing (e.g., Gould 1996), it is possible to do a
microlensing experiment with very faint, unresolved stars in distant
galaxies.  The use of such events has been first discussed for M31 (Crotts
1992; Baillon et al. 1993), and then for M87 in the Virgo
cluster (Gould 1995). Such events may also add new information to
microlensing events towards MCs (Gould 1997; Nakamura \& Nishi 1998; Sumi \&
Honma 2000). A few experiments towards M31 are currently underway (Crotts \&
Tomaney 1996; Ansari et al. 1997; Riffeser et al. 2001; Paulin-Henriksson et
al. 2002; Calchi Novati et al. 2002).

There are a few more approaches other than the pixel lensing, which are
proposed to constrain MACHOs in clusters of galaxies. Walker \& Ireland
(1995) and Tadros, Warren, \& Hewett (1998) considered microlensing of
background quasars behind the Virgo cluster. While the optical depth ($\tau
\sim 10^{-3}$) is larger than the microlensing experiments towards the MCs
($\tau \sim 10^{-7}$), it is not sufficiently large because of the small
number of available background quasars. The problem is also compounded by the
difficulty of distinguishing microlensing-induced quasar variability from
intrinsic mechanisms.  
Lewis \& Ibata (2001) considered to use fluctuation of surface
brightness of galaxies by microlensing to constrain cosmologically
distributed compact objects, and Lewis, Ibata, \& Wyithe (2000) extended this
approach to giant gravitationally lensed arcs in galaxy clusters to constrain
intracluster MACHOs. However, detection of such fluctuation would require a
long observing time of {\it the Hubble Space Telescope} or {\it the Next
Generation Space Telescope}.

Recent developments of advanced observing facilities enable us to do a deep
and/or wide search of microlensing events by ground-based telescopes as well.
The Suprime-Cam installed in the prime focus of the 8.2m Subaru telescope
combines the sensitivity of 8m-class telescopes with a wide field of view of
$30'\times 30'$, and a unique microlensing experiment might be possible by
such a facility.

The Hercules supercluster consists of three rich Abell clusters of galaxies
(A2147, 2151, and 2152) at $z \sim 0.04$, which seems mildly bound
gravitationally, with a total mass of $\sim 8 \times 10^{15} M_\odot$ (Barmby
\& Huchra 1998; Blakeslee et al. 2001). Each cluster seems not completely
stabilized yet showing rather irregular morphologies, and they have
relatively high fraction of spiral galaxies ($\sim$ 50\%). Recently another,
more distant cluster was found just behind A2152 at $z = 0.13$, forming a
cluster-cluster system with a projected separation of only $2.4'$ (=
0.09$h^{-1}$Mpc at $z = 0.04$) (Blakeslee et al. 2001; Blakeslee 2001).
Extremely magnified stars in the background cluster by microlensing of
compact objects in A2152 may be detectable by a deep and wide monitoring of
this region. Here we give an event rate estimate of such phenomena, supposing
a sensitivity of 8m-class telescopes\footnote{The surface brightness of
galaxies is not as high as the sky background in most locations, and hence
brightness of host galaxies would not seriously decrease the sensitivity to
point transient sources.  See \S \ref{section:discussion} in more detail.}
with a reasonable telescope time, and show that the sensitivity is good
enough to do a unique microlensing experiment for a wide range of the compact
object mass.

A similar idea has been studied by Turner \& Umemura (1997), but for stars in
general field galaxies, i.e., not in clusters. They examined only limited
population of source stars and lens mass range, with a simple picture of
point-mass lens.  However, as we will show, caustic-crossing is likely to be
more important when the magnification is extremely large, and simple
point-mass picture does not apply.  We will present formulations by which one
can derive more realistic event rate for a specific observation, taking into
account caustic-crossing, event time scales, and realistic stellar luminosity
function, for a wide range of the lens mass. We will mention that the
cluster-cluster system has some advantages compared with a search made for
general fields.

In \S \ref{section:formulations}, we present general pictures of expected
events and formulations to predict the expected event number.  The prediction
will be made in \S \ref{section:events}, and some discussions will be given
in \S \ref{section:discussion}. Then we will present various astrophysical
implications that will be obtained by this cluster-cluster microlensing
experiment in \S \ref{section:implications}. 
Throughout the paper we use $h \equiv H_0$/(100km/s/Mpc) = 0.7.

\section{Formulations}
\label{section:formulations}
\subsection{General Picture: Point-Mass Lens versus Caustic Crossing}
In microlensing experiments towards the Galactic bulge or MCs, where the
optical depth is much smaller than unity, generally lensing events can be
treated as magnification by a single lens, except for cases where lenses are
forming close binaries.  On the other hand, when optical depth is close to
unity such as microlensing of distant quasars by stars in a galaxy on the
line of sight, the effects of the various microlenses cannot be considered
independently, and complicated caustic networks arise (Wambsganss,
Paczy\'nski, \& Schneider 1990; Narayan \& Bartelmann 1995).  Even if the
optical depth is not as large as unity, caustics still exist in the vicinity
of the center of the individual lenses, by the external shear induced by
nearby point-mass lenses and/or smoothly distributed matter.  Therefore the
effects of shear could be significant when we consider very large
magnification events, even if $\tau \ll 1$.  First we should examine which
picture is appropriate for the case we will consider here. Assuming typical
density profiles in clusters (discussed in more detail later in \S
\ref{section:inputs}), the optical depth $\tau = \Sigma / \Sigma_{\rm crit}$
of the lens cluster A2152 is 0.1--0.2, which is a weighted mean with the
density of stars in the source cluster, if all of the cluster mass is in the
form of compact objects contributing to microlensing. Here, $\Sigma$ is the
surface mass density per unit solid angle, $\Sigma_{\rm crit} = (c^2/4\pi G)
d_S d_L / (d_S - d_L)$ the critical surface density, and $d_S$ and $d_L$ the
distances to the source and lens clusters, respectively.\footnote{ Here,
these distances are angular diameter distances. On the other hand, we will
use $\tilde d_S$ and $\tilde d_L$ for luminosity distances later.  Since the
redshift to the A2152 cluster-cluster system is not large, the cosmological
correction is not important, but we differentiate them for possible
application to more distant systems in the future.}  Therefore microlensing
in this system can marginally be considered as the small optical depth case
$(\tau_{\rm co} \equiv f_{\rm co} \tau \ll 1)$, where $f_{\rm co}$ is the
mass fraction of the compact lenses.  When $f_{\rm co}$ is much smaller than
unity, the small $\tau_{\rm co}$ approximation becomes even better.
Therefore we assume $\tau_{\rm co} \ll 1$ in this paper, and we will consider
only microlensing by one individual microlens. Correction by numerical
studies to this may be necessary in the future in the very central region of
the cluster when $f_{\rm co} \sim 1$.

When $\tau_{\rm co} \ll 1$, microlensing can be
treated as a sum of single point-mass lenses with external shear. The
resulting caustics have the shape of an ``astroid'', with four cusps at an
angular radius $2s \theta_E$ from the location of the lens, where $s$ is the
external shear and $\theta_E$ the Einstein radius
(Chang \& Refsdal 1979, 1984;
Schneider \& Wei\ss 1986; Mao 1992; Kofman et al. 1997).  The
shear is given by the sum of the contributions from nearby compact objects
and smooth mass distribution in the cluster, as $s = s_{\rm co} + s_{\rm
sm}$.  Typically $ s_{\rm co} \sim
\tau_{\rm co}$, and its distribution is given by
\begin{equation}
p(s_{\rm co}|\tau_{\rm co}) = \frac{\tau_{\rm co} s_{\rm co}}
{(\tau_{\rm co}^2 + s_{\rm co}^2)^{3/2}}
\end{equation}
for a random lens distribution (Nityananda \& Ostriker 1984). On the other
hand, assuming a singular isothermal sphere with a softened core, we find
\begin{equation}
s_{\rm sm} = (1-f_{\rm co})
\frac{d_S - d_L}{d_S} \ \frac{4 G M_{\rm cl}}{c^2 R_{\rm cl}} \
\frac{1}{2} \theta^2 (\theta_{\rm core}^2 + \theta^2)^{-\frac{3}{2}} \ ,
\label{eq:isothermal}
\end{equation}
where $M_{\rm cl}$ and $R_{\rm cl}$ are the mass and radius of the lens
cluster, $\theta$ the angular radius from the center of the cluster-cluster
system, and $\theta_{\rm core}$ the angular radius of the core (Narayan \&
Bartelmann 1995). Here we implicitly assumed that all the mass except for the
compact objects is distributed smoothly. The lower limit of $s_{\rm sm}$ is
coming from intracluster gas, which is typically $\sim 10\%$ of the total
cluster mass [i.e., $(1 - f_{\rm co}) \gtrsim 0.1)$].  For a typical
configuration, the weighted mean of this shear by the source star density is
$\sim 0.02$ for the A2152 system if $f_{\rm co} = 0$.

\begin{figure}
\plotone{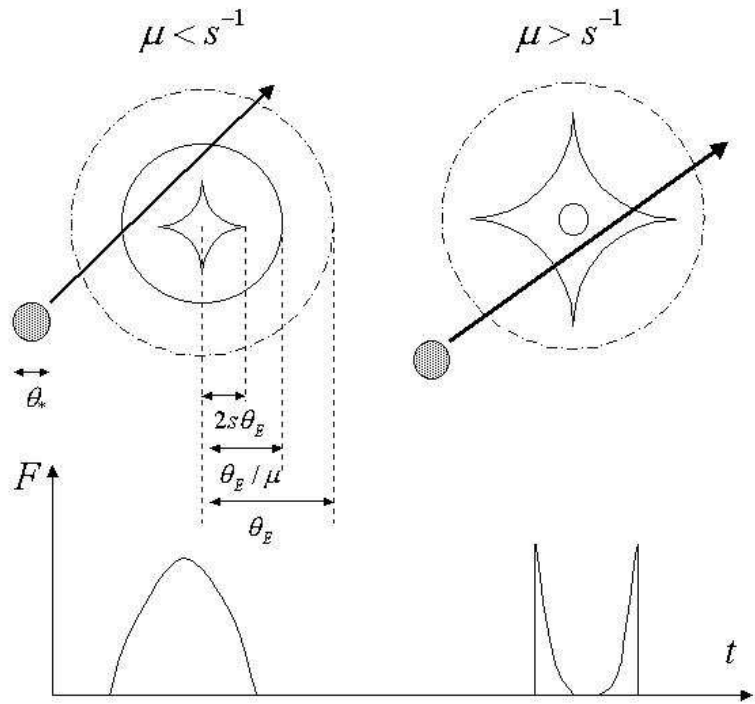}
\caption{ A schematic diagram for the two different cases of microlensing
events: the point-mass lens limit ($\mu<s^{-1}$, left-hand side) and the
caustic-crossing dominant limit ($\mu>s^{-1}$, right-hand side), where $\mu$
is the magnification required for detection and $s$ is the external shear at
the lens location.  The dot-dashed circles are the Einstein radius, and the
solid circles have radii of $\theta_E/\mu$, and a source star with a radius
$\theta_*$ must hit this region for its detection in the point-mass lens
picture. On the other hand, the astroid-shaped curves are caustics, extending
to $\sim 2 s \theta_E$, and a source star must hit this region for its
detection, in the caustic crossing limit.  The corresponding light-curves are
shown in the bottom, where the finite source size effect is assumed to be
negligible. (For the condition of significant finite source size effect, see
text.)  }
\label{fig:schem}
\end{figure}

We should examine whether the caustics around the lens center generated by
the external shear has a significant effect on the microlensing events.  We
use terms of `point-mass lens limit' and `caustic-crossing limit' when the
caustic-crossing effect is negligible or not, respectively.  The appropriate
picture, point-mass lens or caustic crossing, is determined by the value of
the shear $s$ and the magnification required for detection of a microlensed
star, $\mu$. (See a schematic explanation in Fig.  \ref{fig:schem}). When
$\mu \ll s^{-1}$, the impact parameter (the minimum angular separation
between a lens and a source star) required for detection is $\sim
\theta_E/\mu$, which is much bigger than the size of the caustics,
$2s\theta_E$.  In this region outside the caustics, the magnification
behavior is similar to that of an ideal point-mass lens without shear.  On
the other hand, when $\mu \gg s^{-1}$, a star crosses caustics before it
reaches the radius of $\theta_E/\mu$, and hence the strong magnification by
caustic crossing would dominate observable events. As mentioned above, the
external shear is expected to be $\sim$ 0.02--0.2.  Considering the distance
modulus of the source cluster (38.73) and typical sensitivity of 8m-class
telescopes ($m_I \sim 26$ at one hour exposure), we expect that the caustic
crossing is appropriate in most cases, except for the brightest classes of
source stars ($M_I \sim -10$) for which only a magnification of $\mu \sim 10$
is required for detection.  In next subsections we will give formulations to
estimate microlensing event rate for the both limits, and calculate the
expected number of events. Then we examine, for typical events contributing
to the event number, which limit applies in a supposed observation, which
should be dependent on the lens mass considered.

In this paper we consider two modes of observation with total duration of
$T_{\rm obs}$: (1) a consecutive observation during a night, typically
$T_{\rm obs} \sim $ 6 hrs, and (2) a monitoring beyond the time
scale of one night with arbitrary sampling time
interval. The time resolution $t_{\rm res}$ would be minutes for the
observing mode (1) for typical instruments, 
while it is $T_{\rm obs}/N_{\rm sample}$ for the mode
(2), where $N_{\rm sample}$ is the total number of sampling 
(i.e., nights) during the total
observing duration $T_{\rm obs}$. 

\subsection{Point-Mass Lens Limit ($\mu \ll s^{-1}$)}
Let $f_{\rm lim, 0}$ be the flux sensitivity limit with an exposure time
$t_{\rm exp, 0}$, for a given telescope.  We also assume that the sensitivity
limit scales as $f_{\rm lim} = f_{\rm lim, 0} (t_{\rm exp}/t_{\rm exp,
0})^{-1/2}$. First we consider the optimal lensing time scale and
magnification to search for a microlensing event of a source star whose
original luminosity is $L_*$, for the observing mode (1).  The magnification
required for detection at a flux level $f_{\rm lim}$ becomes $\mu = 4 \pi
\tilde d_{\rm S}^2 f_{\rm lim} / L_*$. For this magnification, the impact
parameter between the source star and the lens must be smaller than $\theta <
\theta_E / \mu$ when $\mu \gg 1$, and the time duration of lensing is given
by $t_{\rm lens} = t_{\rm lens, E} \ \mu^{-1}$.  Here, $t_{\rm lens, E} =
\theta_E d_L / V$ is the Einstein-ring crossing time, and $V$ is the
relative transverse velocity between the source and lens projected on the
lens plane.  The lensing duration must be longer than the supposed exposure
time to detect an event, and hence this condition, $t_{\rm lens} \ge t_{\rm
exp}$, results in the limiting magnification $\mu_{\lim}$ required, as:
\begin{equation}
\mu > \mu_{\lim} = \frac{t_{\exp, 0}}{t_{\rm lens, E}} \left( \frac{4 \pi
\tilde d_S^2 f_{\lim, 0}}{L_*} \right)^2 \ .
\end{equation}
For the observing mode (2), the sensitivity is determined by the unit
exposure time $t_u$ during one night (typically $t_u \sim$ 6 hrs), rather
than by the lensing time scale that is longer than $t_u$.  In this case, the
limiting magnification is simply given as $\mu_{\lim} = 4 \pi \tilde d_{\rm
S}^2 f_{\rm lim, u} / L_*$, where $f_{\rm lim, u} = f_{\lim, 0}
(t_u/t_0)^{-1/2}$ is the sensitivity limit for an exposure of the unit time
scale.

We should also examine the finite source size effect. When the impact
parameter $\theta_E / \mu$ becomes smaller than the size of source stars, i.e.,
$r_* > r_{\rm crit} \equiv \theta_E d_S / \mu$, this effect becomes
significant, where $r_*$ is the size of stars.  

To help the reader get a rough image of possible events, we show the values
of representative quantities such as $t_{\rm lens}$, $\mu_{\lim}$, $r_*$ and
$r_{\rm crit}$, for some values of lens mass $m_{\rm co}$ in Table
\ref{table:quantities}.  The observing mode (1) is assumed. For the treatment
of stellar size and other parameters for the cluster-cluster system, see \S
\ref{section:inputs}.  These quantities could be very different for different
source star luminosity, and this means that only stars in a relatively
narrow range of luminosity would contribute to event rate, whose event
time scale $t_{\rm lens}$ is matching the practical observing time scale.

The optical depth for such extremely magnified events is given by $
\tilde{\tau}_{\rm co} = \tau_{\rm co} \mu_{\lim}^{-2} = f_{\rm co} m_{\rm
co}^{-1} \Sigma (\theta) \pi \theta_E^2 \mu_{\lim}^{-2}$, where $m_{\rm co}$ is
the mass of compact lens objects, and $\Sigma (\theta)$ is the surface mass
density of the lens cluster at the angle $\theta$ from the center.  Then the
total number of microlensing events detectable in one snapshot of the source
cluster is given by
\begin{equation}
N_{\rm snapshot} (L_*) dL_* = \int_0^{\theta_{\rm cl, S}} 2 \pi \theta d\theta
S_*(\theta) \phi(L_*)dL_* \tilde \tau_{\rm co}(\theta) \ ,
\end{equation}
for source stars whose luminosity is in a range from $L_*$ to $L_* + dL_*$,
where $S_*$ is the mean surface brightness of galaxies in the source 
cluster\footnote{It should be noted that $S_*$ in
this equation is intrinsic surface brightness, while
observational estimate of $S_*$ is affected by {\it macro}lensing of
the foreground cluster. See \S \ref{section:discussion}.},
$\phi$ the stellar luminosity function (LF) of source stars normalized by the
total stellar luminosity, and $\theta_{\rm cl, S}$ the maximum angular
extension of the source cluster.
Here we implicitly assumed that the two clusters are perfectly
aligned, which is a reasonable approximation for the A2152 system since the
projected angular separation of the centers of the two clusters is comparable
or even smaller than the typical core size of clusters.  For a monitoring
with duration $T_{\rm obs}$, the total expected number of events is then
given by integrating over $L_*$,
\begin{equation}
N_{\rm event} = \int dL_* N_{\rm snapshot}(L_*) \max(1, T_{\rm obs}/t_{\rm
lens}) \ .
\end{equation}   
In practice, the integration range over $L_*$ must be limited,
since the lensing time scale must be reasonable with respect to the finite
time resolution or duration of observation. (Note that $t_{\rm lens}
\propto m_{\rm co} L_*^2$.)
For the observing mode (1), 
we perform this integration when a condition on $t_{\rm lens}$, $5 t_{\rm res}
< t_{\rm lens} < T_{\rm obs}$, is satisfied, 
and for the observing mode (2), we set a condition
$0.5 T_{\rm obs} < t_{\rm lens} < 2 T_{\rm obs}$.

\subsection{Caustic Crossing Limit ($\mu \gg s^{-1}$)}
\label{section:caustic}
First we consider the event rate by lenses with a fixed value of shear $s$,
and then we will integrate it over $s_{\rm co}$ with the probability
distribution $p(s_{\rm co}|\tau_{\rm co})$. It can be shown that the
magnification distribution function $P(\mu)$ becomes asymptotically the same
with that of a point mass lens when $\mu \gg s^{-1}$, and behaves like
$P(\mu) \propto \mu^{-3}$ (Schneider 1987; Kofman et al. 1997). It should be
noted that not only the behavior ($\propto \mu^{-3}$) but also the angular
area in the source plane for magnification larger than a given $\mu$ becomes
the same as those of point mass lens in the limit of $\mu \gg s^{-1}$
(Kofman et al. 1997).  Large magnification events should be dominated by
caustic-crossing rather than cusps, since $P(\mu)$ around the cusps decreases
with increasing $\mu$ as $\propto \mu^{-7/2}$, which is faster than the total
$P(\mu)$ (Mao 1992). For a source crossing the caustics around a lens, the
projected length of caustics on the source plane is $\sim 8s\theta_E$.  Let
$\theta_\mu$ be a characteristic width of the region along the caustics where
magnification is larger than $\mu$. Then we expect that the area of this
region, $\sim 8 s \theta_E \theta_\mu$ should be equal to the equivalent area
for a single point-mass lens, $\pi (\theta_E/\mu)^2$.  Then we obtain
\begin{eqnarray}
\theta_\mu \sim \frac{\pi \theta_E}{8 s \mu^2} \ ,
\label{eq:theta_A}
\end{eqnarray}
which is related with the time duration for this magnification as:
\begin{eqnarray} 
t_\mu = \frac{\theta_\mu d_L}{V} \ .
\label{eq:t_A}
\end{eqnarray}
These relations indicate that the light curve around
the caustic crossing with a constant lens velocity is $\mu \propto
t^{-1/2}$, as is well known for sources inside the caustics, while
magnification suddenly drops when a source crosses and gets outside the
caustics (e.g., Schneider \& Wei\ss 1986).
Since the sensitivity of a telescope also roughly scales as $f_{\lim} \propto
t^{-1/2}$, the signal-to-noise ratio should be roughly constant
against the exposure time in a consecutive monitoring observation.

Therefore all stars brighter than a threshold luminosity can be detected
by microlensing, when they pass a caustic.
For observing mode (1), the magnified flux $f_\mu = \mu L_* / (4
\pi \tilde d_S^2)$ must be greater than the flux limit $f_{\lim} = f_{\lim,
0} (t_\mu/t_0)^{-1/2}$, and hence we obtain
\begin{eqnarray}
L_{*, \min} (s) > 4 \pi \tilde d_S^2 f_{\lim, 0} \sqrt{\frac{8 s V t_0}{\pi
\theta_E d_L}} \ ,
\label{eq:Lmin1}
\end{eqnarray}
which is independent of $t_\mu$ or the exposure time.
In the case of observing mode (2), the
sensitivity limit with an exposure of
the unit observing time $t_u$ should be fainter than the magnified
flux of stars with a lensing time scale of the total observing duration,
i.e., $t_\mu = T_{\rm obs}$. Then we get
\begin{eqnarray}
L_{*, \min}(s) > 4 \pi \tilde d_S^2 f_{\lim, 0} \sqrt{\frac{8 s V t_0 T_{\rm
obs}}{\pi \theta_E d_L t_u}} \ .
\label{eq:Lmin2}
\end{eqnarray}

Ignoring the finite source size effect, the maximum magnification $\mu_{\max}$
that can be observed is determined by the minimum time resolution,
$t_{\rm res}$. Replacing $t_\mu$ by $t_{\rm res}$ in eqs. (\ref{eq:theta_A})
and (\ref{eq:t_A}), we obtain:
\begin{equation}
\mu_{\max} = \sqrt{ \frac{\pi \theta_E d_L}{8  s \ t_{\rm res} V} }  \ .
\end{equation}
On the other hand, $\mu_{\max}$ may be limited by the finite source size
effect. This effect becomes to be visible
when the crossing time of stars,  
$r_* d_L / (V d_S)$, becomes larger than the time resolution
$t_{\rm res}$. This condition can be written as:
\begin{eqnarray}
r_* > r_{\rm crit} \equiv \frac{t_{\rm res} V d_S}{d_L} \ .
\label{eq:finite-size-2}
\end{eqnarray}

These representative quantities, such as $M_{I, \max}$ corresponding
to $L_{*, \min}$, $\mu_{\max}$, $r_*$ corresponding to $L_{*, \min}$,
and $r_{\rm crit}$ are shown for some values of the lens mass,
in Table \ref{table:quantities}. The observing mode (1) is assumed.
It should be noted that $L_{*, \min}$ depends very weakly on the lens mass as
$\propto m_{\rm co}^{1/4}$, and this suggests that this experiment
has a sensitivity in a wide range of the lens mass, as we will see below.
Although the finite source size effect
seems too strong in all cases shown here, it should be noted that here we
used rather small $t_{\rm res} = 0.1$hrs as typically possible time
resolution. We can increase $t_{\rm res}$ to, say, 1 hr for the observing
mode (1) and 1 day for the mode (2). Then we may observe modest effect
of finite source size, which does not seriously decrease detectability of
events, but give important information about the stellar size that is
useful to estimate $m_{\rm co}$.

The expected event number per one source star is calculated
as the expected number of the caustic crossing, whose length is
$\sim 8  s \theta_E$ per microlens, as:
\begin{eqnarray}
N_1(\theta, s_{\rm co}) \ ds_{\rm co} &=& 8 s(\theta) \ \theta_E \frac{V}{d_L} 
n_{\rm co}(\theta)
\ p(s_{\rm co}|\tau_{\rm co}) \ ds_{\rm co} \ T_{\rm obs} \\
&=& \frac{8 s(\theta) \ V
T_{\rm obs} \tau_{\rm co}(\theta) 
\ p(s_{\rm co}|\tau_{\rm co}) \ ds_{\rm co}}{\pi
\theta_E d_L} \ ,
\label{eq:N1}
\end{eqnarray}
for lenses whose $s_{\rm co}$ is in a range from $s_{\rm co}$ to $s_{\rm co}
+ ds_{\rm co}$, where the surface number density of lens is $n_{\rm co} =
\tau_{\rm co} / (\pi \theta_E^2)$.
Note again that $s = s_{\rm co} + s_{\rm
sm}$.  Then the total event number is obtained by integrating over $s_{\rm
co}$ and angular radius from the center of the cluster-cluster system, as:
\begin{eqnarray}
N_{\rm event} &=& \int_0^{\theta_{\rm cl, S}} 2 \pi \theta d\theta
S_*(\theta) \ \int ds_{\rm co} \nonumber \\  &\times& N_*[L_{*, \min}
\{ s(\theta, s_{\rm co}) \} ] \
N_1[\theta, s_{\rm co}] \ ,
\label{eq:Ntot_caustic}
\end{eqnarray}
where $N_*(L_*)$ is the number of stars brighter than $L_*$, i.e.,
\begin{eqnarray}
N_*(L_*) = \int_{L_*}^\infty \phi(L_*') dL_*' \ , 
\end{eqnarray}
which is normalized by
the total stellar luminosity.

\subsection{Image Separations and Time Delays}
In the above formulations, we implicitly assumed that different lensed images
cannot be resolved in observations, and the arrival time difference among
them is negligible compared with observational time scales. Here we check
this point. In the point-mass lens limit, there are two images separated by
$\theta_{\rm sep} \sim 2 \theta_E$, and time delay between the two can be
calculated by the time delay function which is the sum of the geometrical and
gravitational time delays (Narayan \& Bartelmann 1995). To the third order of
$1/\mu$ when $\mu \gg 1$, we find
\begin{equation}
\Delta t = \frac{d_L d_S}{c \ d_{LS}} \ \frac{\theta_E^2}{6 \mu^3} \ .
\end{equation}
In the caustic crossing limit, the separation of newly created
two images when a source just gets inside the caustics 
is given by $\theta_{\rm sep} \sim
\theta_E / \mu s$ ($\theta_{\rm sep} \rightarrow 0$ when the source approaches
to the caustic), and time delay between the two images is given as
\begin{equation} 
\Delta t \sim \frac{d_L d_S}{c \ d_{LS}} \  \frac{\theta_E^2}{\mu^3 s^2} \ ,
\end{equation} 
which has a similar form to the point-mass lens limit
(see, e.g., Schneider \& Wei\ss \ 1986). 
As we will see later, the point-mass lens limit can be applied only for
a small lens mass range of $m_{\rm co} \lesssim 0.1 M_\odot$, and hence
the image separation is practically unresolved by observations.
The factor of $\mu^{-3}$ in $\Delta t$ indicates that the time delay
is much shorter than observing time scales in all the lens mass range
considered in this paper.

\section{Event Rate Estimations}
\label{section:events}
\subsection{Input Parameters}
\label{section:inputs}

We use a cluster total mass $M_{\rm cl} = 10^{15}h^{-1} M_\odot$ for \ the
lensing cluster A2152 by a dynamical mass estimate within $R_{\rm cl, L} = d_L
\theta_{\rm cl, L} = 1.6 h^{-1}$ Mpc (Barmby \& Huchra 1998).  The mass of
the source cluster is uncertain, but it seems more massive than A2152
(Blakeslee 2001). Here we use the same mass and radius with those of A2152.
The velocity dispersion of A2152 is $\sim$700 km/s, and probably it has a
comparable transverse peculiar velocity of bulk motion. Therefore, we use $V$
= 1000 km/s as a plausible value.

The density profiles of lens objects in the lens cluster and source stars in
the source cluster must be specified. For the stellar density profile in the
source cluster, we assume the King profile (King 1962) ($S_* \propto [1 +
(\theta / \theta_{\rm core})^2]^{-1}$), which is the most widely used for the
number distribution of galaxies.  We also assume a core size of $R_{\rm
core} = 0.09 h^{-1}$Mpc in the King profile.  We consider observations in the
$I$ band, and use a mass-to-light ratio of the source cluster to
normalize the surface brightness $S_*$, as $M_{\rm cl} / L_{\rm cl, I} = 180h
(M_\odot/L_{I,\odot})$, which is converted from $M/L_V$ for rich clusters in
Bahcall \& Comerford (2002) using a typical color of cluster galaxies, $V-I =
1.6$. In this paper we use the same King profile also for lens objects in the
lens cluster; this is reasonable when the lens objects are tracing the
stellar mass in the cluster, as expected for intracluster stars.  On the
other hand, the Navarro, Frenk, \& White (1997, hereafter NFW) density
profile may be appropriate for non-interacting dark matter (see also
Bartelmann 1996 for the projected surface density profile of NFW). If we use
the NFW profile [$\rho \propto (c r /R_{\rm cl, L} )^{-1} (1 + c
r/R_{\rm cl, L})^{-2}$] for the lens cluster with a typical
concentration parameter of $c = 5$ (e.g., Allen, Ettori, \& Fabian 2001), we
find that the microlensing probability is reduced by a factor of about 2
compared with the King profile. Finally, we use the singular isothermal
sphere with a softened core (eq. \ref{eq:isothermal}), having the same
core radius with the King profile,
to calculate the shear of smoothly distributed mass.
It is partially for computational simplicity, but also 
a reasonable treatment for intracluster gas.

\begin{figure}
\plotone{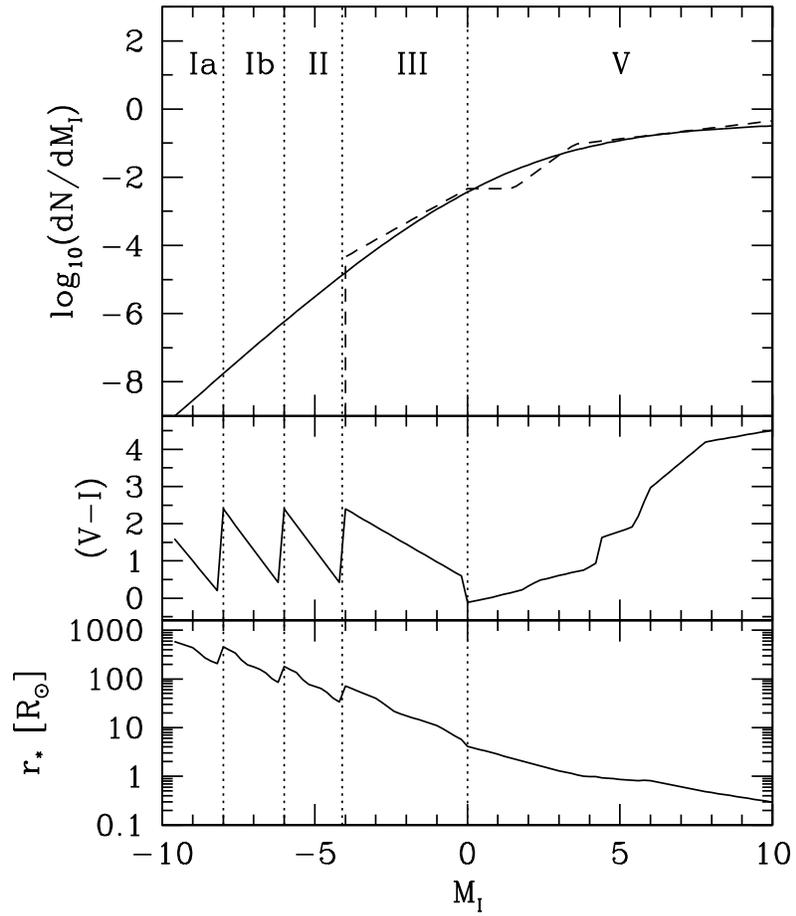}
\caption{ The luminosity function (LF), $(V-I)$ color, and radius of source
stars as a function of $I$ magnitude. Five different classes of stars
[supergiants (Ia and Ib), bright giants (II), giants (III), and main
sequence (V)]
are assumed to be dominant, depending on the magnitude, as shown in the top
panel. The solid line in the top panel is LF in spiral galaxies,
while the dashed line is that in elliptical galaxies having old stellar
populations.
}
\label{fig:lf}
\end{figure}

The LF of source stars must be specified. Here we try two LFs for disk and
elliptical galaxy populations. For disk galaxies, we used the $I$ band LF of
Mamon \& Soneira (1982). For elliptical galaxies or stellar population in
bulges, we use the LF of our Galaxy bulge presented in Terndrup, Frogel, \&
Whitford (1990) showing a cut-off of giant stars brighter than $M_I = -4$,
corresponding to the tip of the red giant branch.  The elliptical/bulge LF
has some structure around $M_I \sim 0$ due to the red clump stars. The
stellar size as a function of stellar luminosity $M_I$ is necessary to check
the finite source size effect. The approximate radius of stars, $r_*$, is
calculated by the bolometric luminosity and the effective temperature ($r_*
\propto L_{*, \rm bol}^{1/2} T_{\rm eff}^{-2}$).  According to Mamon \&
Soneira (1982), we consider five subclasses of stellar populations, i.e., two
supergiant classes (Ia and Ib), bright giants (II), giants (III), and main
sequence (V), which are dominant in different ranges of $M_I$ with an
increasing order, and the borders between these are $M_I \sim -8, -6, -4,$
and 0.  The $(V-I)$ versus $M_I$ relation for these subclasses is also given
in Mamon \& Soneira (1982), from which we infer the spectral type, effective
temperature, and bolometric corrections in Zombeck (1990). The LF, $(V-I)$
color, and stellar radius are shown in Fig. \ref{fig:lf}.

\subsection{Results}
Here we calculate the expected event rate supposing an observation by
Subaru/Suprime-Cam, whose sensitivity is $m_{\lim, 0, I} = 26.0$ (S/N=5) at
$t_{\exp, 0}$= 1 hr. In the point-mass lens approximation, we require a
signal-to-noise of $S/N > 10$ and $S/N > 5$ for the observing modes (1) and
(2), respectively. The higher $S/N$ is required for the observing modes (1)
to assure sufficient $S/N$ to construct a microlensing light curve.  On the
other hand, in the caustic-crossing limit, we require a signal-to-noise of
$S/N > 5$ and $S/N > 3$, respectively. The lower $S/N$ is adopted because the
strong magnification near the caustic-crossing is expected to increase total
effective $S/N$ compared with the point-mass lens limit.  (However, when the
finite source size effect is significant, this may not apply. See below.)

In Figures \ref{fig:f_co_single_1} and \ref{fig:f_co_caustic_1}, we show the
expected limit on $f_{\rm co}$ obtained by an observation using 10 nights,
i.e., 10 times repetition of the observing mode (1) with $T_{\rm obs} = 6$
hrs and $t_{\rm res} = 0.1$ hrs, in the limits of point-mass lens and caustic
crossing, respectively. Figures \ref{fig:f_co_single_2} and
\ref{fig:f_co_caustic_2} are the same, but for the observing mode (2) using
10 nights, with $T_{\rm obs} = 10$ days and $N_{\rm sample} = 10$. In all
these four figures, the limits on $f_{\rm co}$ are shown in the upper panels
as a function of various lens mass. In the lower panels, we plot the mean of
$s \mu$, $r_*$, and $M_I$ for events contributing to $N_{\rm event}$. The
calculation in the point-mass lens or caustic-crossing limits is valid only
when $s \mu \lesssim 1$ or $\gtrsim 1$, respectively.  Here, the limit on
$f_{\rm co}$ is defined as the value where the total expected event number,
$N_{\rm event}$ is unity. It should be noted that $N_{\rm event} \propto
f_{\rm co}$ in the point-mass lens limit, while this relation holds only
approximately in the caustic-crossing limit, since a change of $f_{\rm co}$
would change the shear $s$.  Equations \ref{eq:Lmin1}, \ref{eq:Lmin2},
\ref{eq:N1}, and \ref{eq:Ntot_caustic} indicate that this relation depends on
the shape of LF, and $N_{\rm event} \propto f_{\rm co}$ is exactly valid only
when $N_*(L_*) \propto L_*^{-2}$, which is a roughly correct approximation in
a range of $M_I \lesssim -4$.

The point-mass lens approximation is valid only when $\langle s \mu \rangle
\lesssim 1$, while the caustic-crossing limit is appropriate only when
$\langle s \mu \rangle \gtrsim 1$, where $\langle x \rangle$ represents
a mean of the quantity $x$ over all detectable events.
The transition between the two limits occurs at
$\langle \mu s \rangle \sim 1$, which is at $m_{\rm co} \sim 10^{-5} M_\odot$
in the observing mode (1) and at $m_{\rm co} \sim 0.1 M_\odot$ in the
observing mode (2), respectively. At these transition points, both the limits
are approximately valid and hence the event rate predictions should agree
with each other.  Indeed the two predictions agree at the transition lens
mass scale, providing a support for the validity of our formulations and
calculations. 

Most behavior of the $f_{\rm co}$ limit as a function of $m_{\rm co}$ can be
understood as follows. A trend easily seen is that typical source star
luminosity contributing to event rate becomes smaller with increasing lens
mass. In the point-mass limit, the duration of strong magnification required
for detection is $t_{\rm lens} \propto t_{\rm lens, E} / \mu_{\lim} \propto
m_{\rm co} L_*^2$. Since we are supposing 10 nights duration of observation,
and the detectable time scale is limited by this specific time scale.
Therefore $\langle L_* \rangle \propto m_{\rm co}^{-1/2}$.  In the
caustic-crossing limit, on the other hand, there is the minimum source star
luminosity $L_{*, \min}$ for detectable events, and hence  
$\langle L_* \rangle \propto L_{*, \min} \propto
\theta_E^{-1/2} \propto m_{\rm co}^{-1/4}$. This is why the typical source
star luminosity and limits on $f_{\rm co}$ are less sensitive to $m_{\rm co}$
in the caustic-crossing limit, than the point-mass limit. The event rate in
the point-mass lens limit scales as 
\begin{eqnarray}
R_{\rm event} \propto \frac{dN_*(\langle L_* \rangle)}{d\log L_*} \ \tilde
\tau_{\rm co} \propto \frac{dN_*(\langle L_* \rangle)}{d\log L_*} \ 
 \langle L_* \rangle ^2 \ ,
\end{eqnarray}
per logarithmic stellar luminosity interval. On the other hand, the event
rate in the caustic-crossing limit scales as $\propto N_*(\langle L_* \rangle
) \ \theta_E^{-1} \propto N_*(\langle L_* \rangle ) \ \langle L_* \rangle
^2$. These two have the same dependence on $L_*$, and hence the curves of
$f_{\rm co}$ limit in Figs.
\ref{fig:f_co_single_1}--\ref{fig:f_co_caustic_2} can be understood as
inverted stellar luminosity function per logarithmic interval which is
multiplied by $L_*^2$. Since $dN_*/d\log L_* \propto N_*(L_*) \propto
L_*^{-2}$ at $M_I \lesssim -4$, the limit on $f_{\rm co}$ is roughly constant,
but it becomes weaker with decreasing stellar luminosity at $M_I \gtrsim -4$
because of the change of the luminosity function slope.

The finite source size effect could be significant in the observing mode
(1). (Note that the finite source size effect is not taken into account in
the event rate estimates presented here.)  Although $r_*/r_{\rm crit}$
appears to be much larger than unity in the caustic-crossing limit
(Fig. \ref{fig:f_co_caustic_1}), this partly comes from the use of small
$t_{\rm res}$ = 0.1 hrs in eq. (\ref{eq:finite-size-2}), as mentioned in \S
\ref{section:caustic}.  Since the light curve of caustic crossing is $f
\propto t^{-1/2}$ for a point source, the signal-to-noise ratio does not
change much when we change the monitoring time scale of events. If we choose
a longer time scale, the finite source size effect becomes less significant.
We will be able to increase $t_{\rm res}$ up to a few hours during a night,
and hence the finite source size effect should not severely suppress the
detectable event rate estimated here, especially for larger $m_{\rm co}$.
The finite source size effect is mostly insignificant for the observing mode
(2).

To summarize, these results indicate that monitoring of the cluster-cluster
system using 10 nights of a wide-field 8m-class telescope can probe possible
intracluster compact objects, with a sensitivity to the mass fraction in the
total cluster mass as $f_{\rm co} \sim$ 1--3 \% at $m_{\rm co} \sim
10^{-5}$--$10^{8} M_\odot$ in the observing mode (1), and $f_{\rm co} \sim$
3--10 \% at $m_{\rm co} \sim 10^{-3}$--$10^{10} M_\odot$ in the observing
mode (2). The sensitivity in the observing mode (1) might be somewhat reduced
by the finite source size effect, but we expect that it is not significant.

\section{Discussion}
\label{section:discussion}

A weak point of pixel lensing is that there is a degeneracy of lens
parameters due to the lack of information of the source star luminosity.
Even if we assume the transverse lens velocity as $V \sim 1000$ km/s, the
lens mass cannot be determined if we do not know the source luminosity.  It
is not easy to break this degeneracy, but it might be possible if the color
of microlensing events is measured. Elliptical galaxies have only old stellar
populations and there is a sharp cut off of the stellar luminosity function
at the tip of the red giant branch, where the $V-I$ color of stars becomes
rapidly redder at almost constant $M_I$ on the color-magnitude diagram (e.g.,
Jablonka et al. 1999). Therefore, when a microlensing event with very red
color is observed in an elliptical galaxy, it is very likely that the source
star has an absolute magnitude of $M_I \sim -4$, making the lens mass
estimate possible. When
the finite source size effect is seen in an observed light curve, it also gives
additional information on the apparent stellar size, which can be used to
break the degeneracy (Gould 1997; Sumi \& Honma 2000).

It is important to discriminate the microlensing events from other
astronomical transient sources. In addition to the features generally used in
microlensing searches, i.e., characteristic light curves and achromatic
behavior, there are two expected signatures that are unique for this system:
1) more events are expected for stars in the source cluster $(z = 0.13)$
rather than in the lens cluster $(z = 0.04)$, and 2) the event distribution
is even more concentrated to the cluster center than the matter distribution
in clusters, since the lensing probability is proportional to the product of
the surface densities of the source and lens clusters.

Another interesting possibility that may be useful for discrimination 
and breaking the degeneracy is
repetition of caustic crossings. When a source star crosses the
astroid-shaped caustics of a lens, typically two, and sometimes even more
caustic-crossings are expected. The time interval of repetition is roughly
given as 
\begin{equation}
t_{\rm rep} \sim \frac{2 s \theta_E d_L}{V} =
3.0 \times 10^6 \left( \frac{m_{\rm co}}{M_\odot} \right)^{1/2}
\left( \frac{s}{0.01} \right) \ \rm s \ .
\end{equation}
Therefore, repetition of caustic-crossings is expected during the
observing duration $T_{\rm obs} \sim$ 10 days, if the lens mass
is smaller than $m_{\rm co} \lesssim 0.1 M_\odot$. Such repeating
events at the same location in a host galaxy with characteristic
light-curves of caustic crossing would be a good evidence for microlensing,
and also provide independent information of $\theta_E$, and hence, the
lens mass.

In the microlensing experiments towards the Magellanic Clouds, the
self-lensing event rate by stars in MCs is comparable to those by lenses in
the Galactic halo. This has been one of the major causes of the controversial
interpretations of the microlensing events to MCs (e.g., Jetzer, Mancini, \&
Scarpetta 2002).  The situation is similar also for the pixel lensing
experiments towards the M31 galaxy (Paulin-Henriksson et al. 2002) or M87 in
the Virgo cluster (Gould 1995).  However, in the cluster-cluster system, the
lens cluster is located far from both the source system and the observer, and
the lensing events by compact objects in the lens cluster should dominate
those in the source cluster, because of the larger Einstein radius. In fact,
the optical depth to the self-lensing does not depend on the distance, but is
simply $\tau \sim (\upsilon_{\rm vir}/c)^2 \sim 10^{-7}$--$10^{-6}$ which is
much smaller than that of the cluster-cluster system, where $\upsilon_{\rm
vir} \sim 200$ km/s is the virial velocity of the stellar system.

We should examine that the microlensing event rate in the cluster-cluster
system are sufficiently higher than contaminating events by stars or lenses
in the field outside the source and lens clusters. First we consider the case
that the stars in the source cluster are lensed by compact objects outside
the lens cluster (i.e., in the field).  A comparison can be made in terms of
the optical depth; taking the matter density of the universe to be $\Omega_M
= 0.3$, we found that the field optical depth to the distance $d_S$ is $4.2
\times 10^{-3}$, which is more than 10 times smaller than that of the lens
cluster, 0.1--0.2, which is a weighted mean with the source star surface
density.  Secondly, we consider the case where stars in field galaxies behind
the lens cluster are lensed by compact objects in the lens cluster. Assuming
a constant stellar to total mass ratio for clusters and fields, about a third
of the stellar mass of the source cluster is included in the field galaxies
in the cone made by the observer and a projected surface area of the source
cluster. For these field stars, mean optical depth of the lens cluster is
$\sim 0.03$ assuming the King profile. Combining these factors, it can be
concluded that the events by source stars in the field is more than 10
times less frequent than those by stars in the source cluster.

Apart from microlensing, the mean steady flux of stars and galaxies in the
background cluster are also magnified by macrolensing, i.e., gravitational
lensing effect of the mass distribution on the cluster scale. This effect
makes less luminous stars detectable than we considered by the above
formulations, and hence increasing the event rate. According to the
modeling of Blakeslee et al. (2001), we expect that macrolensing
magnification is greater than a factor of 1.5 in the central 
30'' radius region of A2152, for background sources at $z \sim 0.13$.
The magnification becomes more than 10 within the 10'' radius.
These regions are not significant compared with the total cluster field,
but somewhat comparable with the core regions of these clusters. Since
a considerable fraction of projected mass is included in the core regions,
this effect could be significant and may be observed as even stronger
concentration of events to the core regions, than expected from the
product of surface densities of the source and lens clusters. On the
other hand, it should also be noted that the macrolensing effect must
be corrected when one estimates the mean galaxy surface
brightness, $S_*$, from observed galaxy distributions.

When a microlensing event occurred in a very high surface-brightness 
region of a galaxy, the sensitivity to transient point sources could be
effectively reduced. However, generally ground-based observations are
limited by the sky background rather than surface brightness of galaxies. Assuming a sky
background of 19.5 mag arcsec$^{-2}$ in the $I$ band for a moderately dark
sky at Mauna Kea, we estimated about 74\% of the galactic light of the source
cluster at $z = 0.13$
is coming from a region whose surface brightness is lower than the sky background, by
using the local galaxy luminosity function, the mean luminosity-size relation
of galaxies, and surface brightness profiles presented in Totani \& Yoshii (2000). Here, we
assumed a condition of 1 arcsec seeing, and the morphological type mix is
taken to be 70\% for the ellipticals and 30\% for spirals.
Therefore, effective reduction of the sensitivity due to high surface brightness of
host galaxies is not a serious problem.

\section{Astrophysical Implications}
\label{section:implications}

The sensitivity of $f_{\rm co} \sim$ a few percent in a range of $m_{\rm co}
\sim 10^{-5}$--$10^{10} M_\odot$ is sufficiently good as a probe for the nature
of dark matter; we could detect or reject any compact objects in this mass
range as the dominant component of the dark matter in galaxy clusters. 
It should be noted that
this mass range fills up the ``desert'' of the constraints on $\Omega_M$ in
the form of compact objects: $m_{\rm co} \sim 10$--$10^5 M_\odot$, 
which has hardly been constrained by
past observations. Microlensing searches in nearby galaxies or quasar
variability have constrained at $m_{\rm co} \lesssim 10 M_\odot$,
while millilens searches for
radio quasars or echos of gamma-ray bursts  constrained at 
$m_{\rm co} \gtrsim 10^5 M_\odot$ (Narayan
\& Bartelmann 1995; Nemiroff et al. 2001; Wilkinson et al. 2001, and
see Wambsganss 2002 for the latest review). 

Hawkins (1993, 1996) claimed that variability seen in
high-$z$ quasars is due to the microlensing action of Jupiter-mass compact
objects distributed cosmologically, whose density is enough to explain a
significant fraction of the dark matter. However, this claim has been
questioned by a number of authors (e.g., Baganoff \& Malkan 1995; Alexander
1995). Some observations of strongly lensed quasars have been used to
exclude this possibility (Schmidt \& Wambsganss 1998; Wyithe, Webster,
\& Turner 2000), although it also depends on assumed 
quasar sizes. The latest data of EROS project also seem to have excluded
this possibility (Lasserre et al. 2000). Anyway, this experiment would 
provide another independent test for this controversial claim.

If MACHOs exist in the intracluster space with a similar mass fraction
($\sim$ 20\%) suggested by the MACHO collaboration (Alcock et al. 2001), the
cluster-cluster microlensing search should find about 1--10 events.  However,
it should be noted that the mass-to-light ratio of clusters is much larger
than that of the Galactic halo.  If the abundance of MACHOs scales with
luminous matter, we expect that the intracluster MACHO mass fraction is much
smaller than in the Galaxy.  Some observations suggest that MACHOs may be
white dwarfs (Ibata et al. 2000; Oppenheimer et al. 2001), but if 20\% of the
total cluster mass is in the form of white dwarfs, and the matter content in
the cluster system is the same as that of the whole universe, almost all of
the cosmic baryons predicted by the big-bang nucleosynthesis [$\Omega_B \sim
0.02 h^{-2} = 0.07 h^{-2} \Omega_M (\Omega_M/0.3)^{-1}$,
Burles \& Tytler (1998)] must be locked up in
white dwarfs. Such a case would easily violate the constraint coming from the
cosmic background radiation (CBR) in optical and infrared bands (Madau \&
Pozzetti 2000).

On the other hand, there are a few recent reports by independent groups for
detections of the near-infrared CBR (Matsumoto 2000;
Cambr\'esy et al. 2001; Wright 2001),
and the reported flux is by a factor of a few higher than the flux
integration of galaxy counts in the same band, which is difficult to explain
by normal galactic light even if the incompleteness of galaxy surveys and the
cosmological surface brightness dimming of galaxies are taken into account
(Totani et al. 2001). Though the discrepancy may be solved if there are some
systematic errors in processes of diffuse CBR measurement, e.g., subtraction
of the zodiacal light (Wright \& Johnson 2001), it is also possible that
the CBR excess is due to exotic extragalactic sources that are very
different from normal galaxies.
If white dwarfs that were formed at very high redshift
are responsible for this excess ($I \sim 30 \rm nW \ m^{-2} sr^{-1}$), the
mass density of white dwarfs and their progenitors would be about 2\% and
10\% of the nucleosynthetic baryons, respectively, assuming that 80\% of
baryons in progenitors is returned into interstellar space (Madau \& Pozzetti
2000). Therefore, a plausible fraction of such white dwarfs in the total
cluster mass is only $\sim$ 0.2\%. It is not impossible, but rather difficult
to detect these objects by the cluster-cluster microlensing experiment,
unless a large number of nights are available.

Another possible source of this excess of CBR is the first generation stars
at redshift $z \sim 10$, whose UV and optical light is redshifted to the near
infrared band (Santos, Bromm, \& Kamionkowski 2002; Schneider et al. 2002;
Salvaterra \& Ferrara 2002).
Recent theoretical studies on the formation of primordial stars strongly
indicate that they are very massive ($\gtrsim 100 M_\odot$), and a majority
of them might eventually evolve into massive black holes without ejection of
any amount of heavy elements. Then, a major episode of the first-generation
star formation is possible before the interstellar matter is polluted by
metals and normal star formation begins (Schneider et al. 2002).  Assuming a
conversion efficiency from the rest mass to radiation energy that is similar
to normal stars, a mass comparable with the present-day stars [$\Omega_* \sim
0.0024 h^{-1}$, Fukugita, Hogan, \& Peebles (1998)] must have been
locked in the first generation stars and then their remnant black holes with
$M \gtrsim 100 M_\odot$, to explain the CBR excess.  If such black holes are
diffusely distributed in intracluster medium, the cluster-cluster
microlensing search might detect them.

It is expected that there is a diffuse population of intracluster stars that
are stripped from galaxies by interactions with other galaxies or
intracluster gas.  Observations of diffuse optical light, intracluster
planetary nebulae, and red giant stars indicate that the amount of stellar
light from such intracluster stars is 10--50\% of the total light from
galaxies in clusters, though these estimates are still highly uncertain
(e.g., V\'ilchez-G\'omez, Pell\'o, \& Sanahuja 1994; Gonzalez et al. 2000;
Arnaboldi et al. 2002; Durrell et al. 2002; Okamura et al. 2002). As shown
above, about one percent of $\Omega_M$ is locked in stars in the universe,
and this fraction is probably even higher in clusters of galaxies because of
larger fraction of elliptical galaxies. Therefore, if the amount of
intracluster stars is comparable with that of stars in member galaxies, the
microlensing search of the Hercules supercluster might detect them, providing
a completely independent information for intracluster stars.

The author is deeply indebted to B. Paczy\'nski for many inspiring
conversations.  He would also like to thank M. Chiba, E. Komatsu, S. Mao,
J. Ostriker, T. Sumi, J. Wambsganss, and M. Kubota for useful information and
discussions.  He has been financially supported in part by the JSPS
Postdoctoral Fellowship for Research Abroad.


\begin{table}
\caption{Quantities for Some Representative Cases}
\begin{tabular}{cccccccccccc}
\hline \hline 

& & \multicolumn{5}{c}{Point-Mass Lens Limit} & &
\multicolumn{4}{c}{Caustic Crossing Limit}  \\ 

\cline{3-7} \cline{9-12}

$m_{\rm co}[M_\odot]$ & $\theta_E [\mu{\rm as}]$ & $M_I$ & $t_{\rm
lens}$[hrs] & $\mu_{\lim}$ & $r_* [R_\odot]$ & $r_{\rm crit} [R_\odot]$ & &
$M_{I, \max}$ & $\mu_{\max}$ & $r_* [R_\odot]$ & $r_{\rm crit} [R_\odot]$ \\

\hline

$10^{-6} $ & $5.8 \times 10^{-3}$ & $-10$ & 11 & 3.7 & 400 & 190 &&
$-8.75$ & $1.3 \times 10^2$ & 200 & 1.7\\ 

$10^{-3} $ & $1.8 \times 10^{-1}$ & $-7$ & 44 & 29 & 150 & 770 &&
$-6.9$ & $7.1 \times 10^2$ & 150 & 1.7 \\ 

$1 $ & 5.8 & $-3$ & 29 & $1.4 \times 10^3$ & 40 & 500 &&
$-5.0$ & $4.0 \times 10^3$ & 50 & 1.7 \\

$10^{3} $ & $1.8 \times 10^2$ & 1 & 19 & $6.8 \times 10^4$ & 3 & 330 &&
$-3.1$ & $2.2 \times 10^4$ & 30 & 1.7 \\

$10^{6} $ & $5.8 \times 10^3$ & 5 & 12 & $3.3 \times 10^6$ & 0.9 & 220 &&
$-1.3$ & $1.3 \times 10^5$ & 10 & 1.7 \\

$10^{9} $ & $1.8 \times 10^5$ & 9 & 7.6 & $1.7 \times 10^8$ & 0.4 & 140 &&
$0.6$ & $6.8 \times 10^5$ & 4 & 1.7 \\

\hline
\hline
\tablecomments{Sensitivity limit is assumed to be $m_{I, \lim} = 26$
at an exposure time $t_0 = 1$ hr. The observing mode (1) is assumed.
The external shear $s = 0.01$ and
the minimum time resolution $t_{\rm res} = 0.1$ hrs are used
for the caustic crossing limit. Col. (1): the lens mass. Col. (2):
the Einstein radius. Col. (3): Absolute magnitude of source stars.
Col. (4): optimal lensing time scale for detection for $M_I$ given in
the third column. Col. (5): the minimum magnification required for detection.
Col. (6): radius of source stars corresponding to $M_I$. Col. (7):
the critical radius at which the finite source size effect becomes 
significant. Col. (8): the faintest absolute luminosity of source stars
that are detectable. Col. (9): the maximum magnification possible for
the assumed resolution time $t_{\rm res}$. Cols. (10) and (11): the same
as cols. (6) and (7), respectively, but for the caustic crossing limit.}

\end{tabular}

\label{table:quantities}
\end{table}


\epsscale{0.7}

\begin{figure}
\plotone{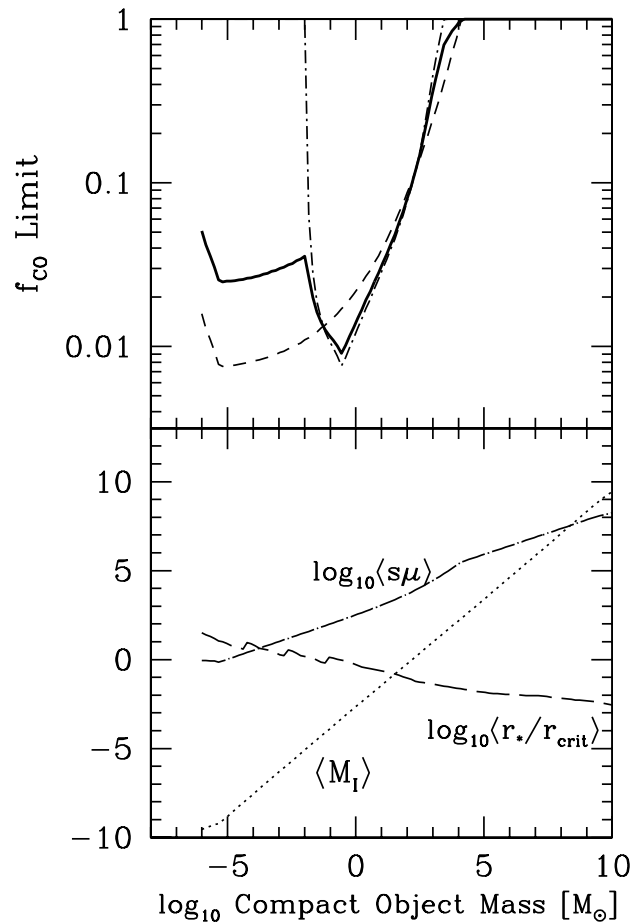}
\caption{ The sensitivity to the mass fraction $f_{\rm co}$ of compact
objects in the total cluster mass, in the limit of point-mass lens
approximation (valid only when $s \mu \lesssim 1$), by 10 times repetition of
consecutive monitoring during a night (6 hrs) with observing mode (1).  Top
panel: the limit on $f_{\rm co}$ as a function of the lens mass, assuming the
stellar luminosity function in spiral galaxies (dashed line), in elliptical
galaxies (dot-dashed line), and the weighted mean with relative proportions
of 30\% for spirals and 70\% for ellipticals (solid line). Bottom panel: the
mean values of original absolute $I$ magnitude of source stars ($M_I$), 
product of magnification and shear ($s \mu$), and the ratio of
the stellar size to the critical size for finite source size effect
($r_*/r_{\rm crit}$). }
\label{fig:f_co_single_1}
\end{figure}

\begin{figure}
\plotone{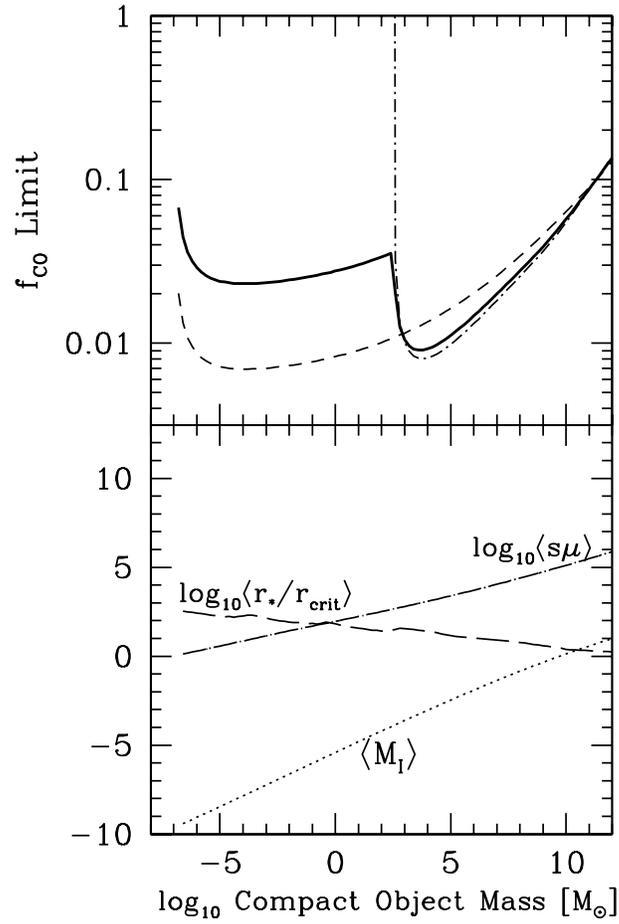}
\caption{ The same as Fig. \ref{fig:f_co_single_1}, by 10 times repetition of
consecutive monitoring during a night (6 hrs) with observing mode (1), but in
the limit of the caustic-crossing approximation, which is valid only when $s 
\mu \gtrsim 1$.  }
\label{fig:f_co_caustic_1}
\end{figure}

\begin{figure}
\plotone{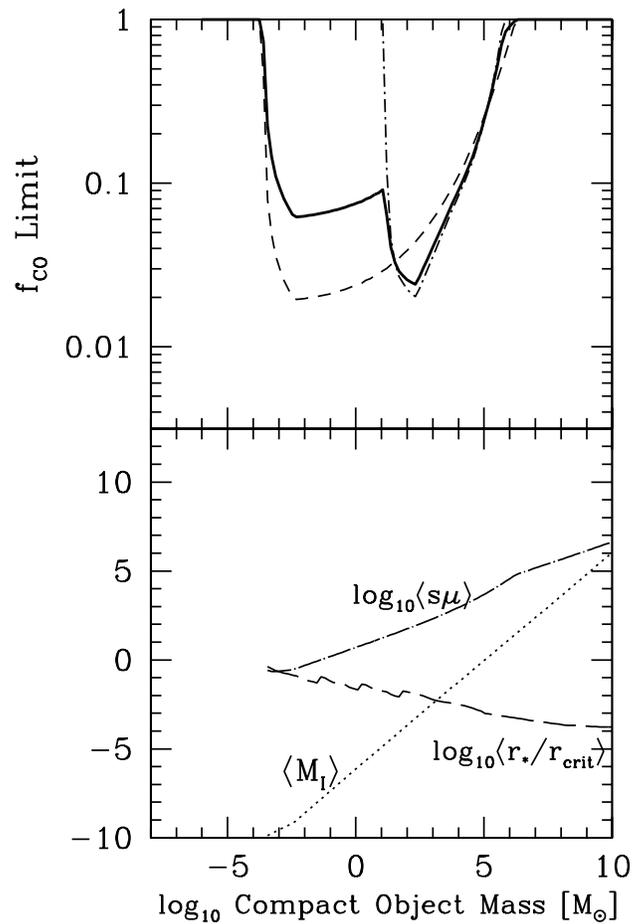}
\caption{ The same as Fig. \ref{fig:f_co_single_1}, but by the observing
mode (2) with $T_{\rm obs} = 10$ days and $N_{\rm sample} = 10$.
This figure is assuming the single lens limit, which is valid
only when $s \mu \lesssim 1$. 
}
\label{fig:f_co_single_2}
\end{figure}

\begin{figure}
\plotone{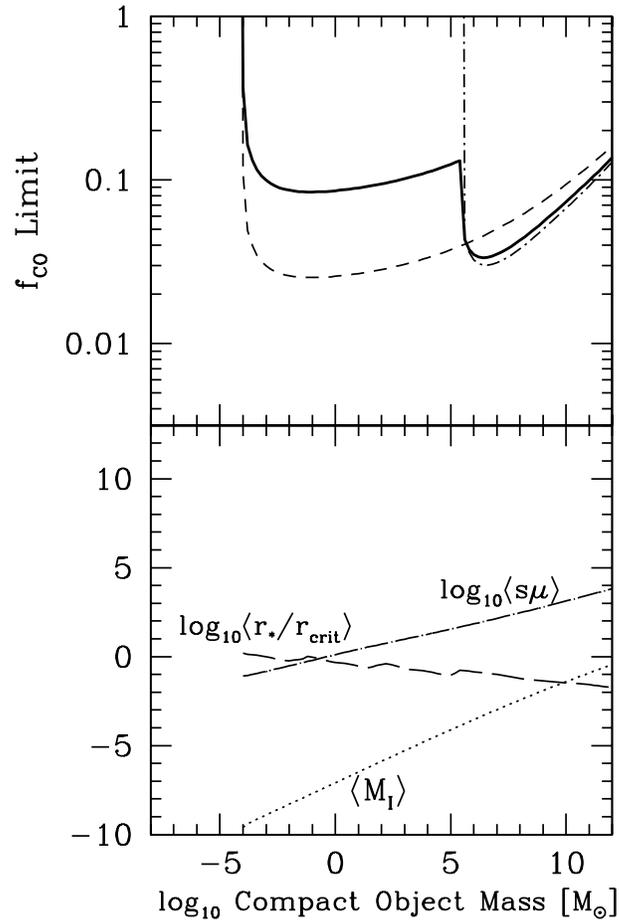}
\caption{ The same as Fig. \ref{fig:f_co_single_1}, but by the observing
mode (2) with $T_{\rm obs} = 10$ days and $N_{\rm sample} = 10$, and 
in the limit of the
caustic-crossing approximation, which is valid only when $s \mu \gtrsim 1$.
}
\label{fig:f_co_caustic_2}
\end{figure}

\end{document}